\documentstyle[preprint,aps]{revtex}

\begin{document}
\draft
\title{Nontrivial resonance of nanoscale uniaxial magnets
to alternating field}

\author{Seiji MIYASHITA, Keiji SAITO}

\address{
Department of Earth and Space Science, Faculty of Science, \\
Osaka University, Toyonaka, Osaka 560 }

\author{Hans DE RAEDT}

\address{Institute for Theoretical Physics and Materials Science Centre,\\
University of Groningen, Nijenborgh 4, NL-9747 AG Groningen, The Netherlands}
\date{23 June 1997}
\maketitle

\begin{abstract}%
How nanoscale uniaxial magnets respond to an alternating field is
studied by direct numerical calculation. A nontrivial oscillation
of the magnetization is found, which is analyzed in terms of
the non-adiabatic transition due to the time dependent field.
A new method to estimate the tunneling gap of the magnet is proposed.
\end{abstract}

\narrowtext

The quantum dynamical behavior of nanoscale magnets have
attracted interests both theoretically
\cite{thmy1,thmy2,thmy3,th4,th5,th6,th7,th8}
and experimentally. \cite{exp1,exp2,exp3,exp4,exp5}
In particular, the step-like magnetization process of Mn$_{12}$-Ac is
a peculiar phenomena, due to quantum dynamics,
for which various explanations have been proposed.
We have also proposed a mechanism in terms of successive
Landau-Zener-St\"uckelberg(LZS)
transition.\cite{Landau,Zener,St}
We pointed out that the non-adiabatic transition between
nearly degenerate levels gives essential the mechanism for the steps and
that
the velocity of the changing field, $c$, is an important parameter because
the transition probability is very sensitive to $c$.

For the LZS transition, so far mainly a linear time dependence of the
field, $H(t)=ct$, has been considered, although an effect of oscillating field
has been studied in a very different context. \cite{Pd}
In the present letter, we study the quantum mechanical response
to an alternating field for a simple model of a uniaxial magnet.
Here we study the transverse-field Ising model:
\begin{equation}
{\cal H}=-J\sum_{<i,j>}\sigma_i^z\sigma_j^z-\Gamma\sum_{i=1}^L\sigma_i^x
-H(t)\sum_{i=1}^L\sigma_i^z,
\label{eqmodel}
\end{equation}
where
\begin{equation}
H(t)=H_0\cos\omega t.
\label{eqht}
\end{equation}
Throughout this letter we take $J$ as a unit of energy and put it unity.
Here we show only results for a system
of four spins ($L=4$) subject to periodic boundary conditions and
$\Gamma=0.5$.
For other choices of model parameters, we find qualitatively similar behavior.
In Fig. 1 the energy levels of the model are shown as a function
of the field $H=H(t)$.
Only the 8 lowest states, some of which are degenerate, are shown.
When the energy gap at $H=0$ is small,
the lowest two levels are located far below the other levels.
When we take the initial state to be the ground state,
the system can be regarded as a two level system
as far as $H_0$ is small and does not cause the second
scattering to the higher levels.
Successive non-adiabatic transitions to higher levels
have been studied in Ref. \cite{thmy3}.

If $H_0$ is very small, we may use the Kubo formula to
study the linear response,
\begin{equation}
\langle M_z(t)\rangle={\rm Re}(\chi(\omega)H_0e^{i\omega t}),
\end{equation}
where $\chi(\omega)$ is the dynamical susceptibility
\begin{equation}
\chi(\omega)=-\lim_{\epsilon\rightarrow +0}{i\over\hbar}
\int_0^{\infty}{\rm Tr}[M_z,M_z(t)]
e^{-\beta{\cal H}_0-i\omega t -\epsilon t}dt/Z,
\end{equation}
which is given at zero temperature,
namely $\beta\rightarrow\infty$, as
\begin{equation}
\chi(\omega)=\sum_{\ell}|\langle G|M_z|\ell\rangle|^2
\left[{\cal P}
{2\hbar\omega\over (E_{\ell}-E_G)^2-(\hbar\omega)^2}
-i\pi\delta(E_{\ell}-E_G-\hbar\omega)
\right]
\end{equation}
where
${\cal H}_0$ is ${\cal H}$ with $H_0=0$, $M_z=\sum_i\sigma_i^z$ and
$Z={\rm Tr}e^{-\beta {\cal H}_0}$.
$E_G$ and $E_{\ell}$ are the energy of the ground state and
the $\ell$-th excited state, respectively.
In this formula the Zeeman term $H(t)M_z$ is treated as a perturbation
and relevant frequencies are only those due to the energy gaps at $H=0$.
On the other hand, in the present paper we are interested in the
phenomena due to the non-adiabatic transition where $H(t)M_z$ can not be
treated as a perturbation. Thus even when we say $H_0$ is small,
$H_0$ is still $O(1)$ and not small enough to be treated as a perturbation.

The probability for staying in the ground state
when the field changes the sign is given as \cite{thmy2}
\begin{equation}
p=1-\exp\left( -{\pi(\Delta E)^2\over 4c|M_0|}\right),
\label{eqp}
\end{equation}
where $c$ is the velocity which is given by
\begin{equation}
c=\left| {{\rm d}\over {\rm d}t}H(t)\right|_{H(t)=0}=H_0\omega
\label{eq}
\end{equation}
and $M_0$ is the ground state magnetization near $H=0$,
\begin{equation}
M_0=\lim_{H\rightarrow 0}M(H).
\label{eqm0}
\end{equation}
For $\Gamma=0.5$,
the energy gap at $H=0$ between the ground state and the first
excited state, $\Delta E$, is 0.03549 and $|M_0|\sim L=4$.

The time evolution of system is given by
\begin{equation}
|t\rangle=e^{-i\int_0^t {\cal H}(s){\rm d}s/\hbar}|0\rangle,
\label{eqev}
\end{equation}
where $|0\rangle$ is an initial state which is chosen to be the
ground state of the model for $H=H(0)$ and the exponential denotes
the time-ordered exponential.
We solve Eq.(\ref{eqev}) making
use of the 4-th order decomposition proposed by Suzuki.\cite{Suzuki1,Suzuki2}
Hereafter we put $\hbar=1$ for simplicity.

As has been shown in the previous studies, Eq.(\ref{eqp}) is
confirmed by the simulation results. From Eq.(\ref{eqp}), $p=0.0062$
for $\omega=0.2$ and $H_0=0.2$.
In the simulation we calculate the overlap
between the ground state and the first
excited state \cite{thmy1}
\begin{equation}
x(t)=|\langle G(t)|t\rangle|^2,
\label{eqx}
\end{equation}
where $|G(t)\rangle$ is the ground state for $H=H(t)$.
After a half period, $t=\pi/\omega$, we find
$x(t=\pi/\omega)=|\langle G(t)|t\rangle|^2=0.0063$,
which confirms the LZS prediction.

In Fig. 2 we show the time dependence of the magnetization,
\begin{equation}
m(t)=\langle t|\sum_i\sigma_i^z|t\rangle,
\label{eqm}
\end{equation}
and observe a gradual relaxation due to the successive
non-adiabatic transitions.
When we continue the simulation, a sinusoidal motion is found
as shown in Fig. 3,
\begin{equation}
m(t)\sim \cos(\Omega t),
\label{eqmt}
\end{equation}
where $x(t)$ and $H(t)$ are also shown.
The period of this sinusoidal motion does not correspond to an
eigenfrequency of the system nor to the period of the external field.
Actually when we change the amplitude of the field $H_0$
the period of the magnetization changes as shown in Fig. 4(a).
The dependence of the period on $\omega$ is also shown in Fig. 4(b).

Although the dependence of $\Omega$ on $H_0$ in Fig. 4 seems
irregular, we find a rather regular dependence when we plot
the frequency $\Omega$ as a function of $H_0$,  shown in
Fig. 5.

Let us study the dependence of $\Omega$ on $H_0$ and $\omega$.
The time-evolution during the field changes from $H_0$ to $-H_0$
is given by
\begin{equation}
X=\exp\left( -i\int_0^{\pi/\omega}{\cal H}(s){\rm d}s\right),
\label{eqX}
\end{equation}
which is 2$\times$2 unitary matrix
as far as $H_0$ is small and only the lowest two states takes
dominant role.
Here we take the ground state $|G\rangle$ and the first excited
state $|1\rangle$ for the initial state as the basis.
Because the probability $p$ is known as Eq.(\ref{eqp}),
we take the following form of $X$
\begin{equation}
X=\left(\matrix{\sqrt{1-p}, & e^{i(-\theta+\phi)}\sqrt{p}\cr
                e^{i\theta}\sqrt{p}& -e^{i\phi}\sqrt{1-p}\cr}
\right),
\label{eqX2}
\end{equation}
where $\theta$ and $\phi$ are unknown phases which depend
on  $H_0$ and $\omega$.
After $t=\pi/\omega$, an inverse process is taken.
If we change the sign of $z$-component of the spins,\footnote{
For this change we can use the unitary transformation
$(\sigma_x,\sigma_y,\sigma_z)\rightarrow(\sigma_x,-\sigma_y,-\sigma_z)$
for all sites.}
the time-evolution during the field changes from $-H_0$ to $H_0$
is identical to $X$, because the motion of the Hamiltonian is
identical.
Thus we only have to change the basis.
The ground state $|G'\rangle$ and the first excited state $|1'\rangle$ for
$t=\pi/\omega$ is generally
expressed as a linear combination of $|G\rangle$ and $|1\rangle$:
$|G'\rangle=a|G\rangle+b|1\rangle$ and
$|1'\rangle=c|G\rangle+d|1\rangle$.
Let the transformation matrix be $Q$.
Thus the second half time-evolution $X'$ is expressed as
\begin{equation}
X'
=\exp\left( -i\int_{\pi/\omega}^{2\pi/\omega}{\cal H}(s){\rm d}s\right),
=Q^{-1}XQ.
\label{eqX2p}
\end{equation}
When the scattering region of $H(t)$ is very narrow, which is the
present case as shown in Fig. 1, we may take
\begin{equation}
Q=\left(
\matrix{0&1\cr 1&0\cr}
\right).
\label{eqQQ}
\end{equation}
Combining Eqs.(\ref{eqX2}),(\ref{eqX2p}) and (\ref{eqQQ}),
the time-evolution operator for one period is given as
$$ {\cal L}=Q^{-1}XQX $$
\begin{equation}
=\left(\matrix{
e^{2i\theta}p+(1-p)e^{i\phi},
& (e^{i\theta+i\phi}-e^{-i\theta+2i\phi})\sqrt{p(1-p)}\cr
 (e^{i\theta}-e^{-i\theta+i\phi})\sqrt{p(1-p)},&
e^{-2i\theta}p+(1-p)e^{i\phi} \cr }
\right).
\label{eqL}
\end{equation}
The eigenvalues, $\lambda_{\pm}$, of ${\cal L}$ are given by
\begin{equation}
\lambda_{\pm}=(q\pm i\sqrt{1-q^2})e^{i\phi}
\label{eqlembda}
\end{equation}
where
\begin{equation}
q=1-p+p\cos(\alpha), \ \alpha=2\theta-\phi.
\label{eqq}
\end{equation}
Now we put
\begin{equation}
\lambda_{\pm}
\equiv e^{\pm i({2\pi\over \omega})({\Omega\over2})+i\phi},
\label{eqlembda2}
\end{equation}
where $2\pi/\omega$ is the period of the field and we
take $\Omega$ to give the frequency in Eq.(\ref{eqmt}).
Here
\begin{equation}
\tan\left({\pi\Omega\over \omega}\right)={\sqrt{1-q^2}\over q}
={\sqrt{2p(1-\cos\alpha)-p^2(1-\cos\alpha)^2}
\over 1-p(1-\cos\alpha)}.
\label{eqtanO}
\end{equation}
When $p\ll 1$,
\begin{equation}
{\pi\Omega\over \omega}\simeq \sqrt{2p(1-\cos\alpha)}.
\label{eqtanO2}
\end{equation}
The evolution of the wave function is generally given by
$${\cal L}^n\left(\matrix{1\cr 0\cr}\right)=
{\cal L}^n\left[c_+\left(\matrix{x_+\cr y_+\cr}\right)
+c_-\left(\matrix{x_-\cr y_-\cr}\right) \right]$$
\begin{equation}
=c_+e^{i{\pi\Omega\over \omega}n}\left(\matrix{x_+\cr y_+\cr}\right)+
 c_-e^{-i{\pi\Omega\over \omega}n}\left(\matrix{x_-\cr y_-\cr}\right),
\label{eq20}
\end{equation}
where $^t(x_{\pm},y_{\pm})$ are the eigenvectors and $c_{\pm}$ are
coefficients.
Thus the probability being in the ground state after $n$ periods,
$x(2\pi n/\omega)$, is given by
\begin{eqnarray}
x(2\pi n/\omega) &=&|c_+x_+e^{i{\pi\Omega\over \omega}n}+
c_-x_-e^{-i{\pi\Omega\over \omega}n}|^2 \\
\nonumber
 &=& a+b\cos({2\pi\Omega\over \omega}n+\gamma).
\label{eqxnt}
\end{eqnarray}
When $p\ll 1$, $x_+\simeq x_-\simeq y_+\simeq -y_-\simeq 1/\sqrt 2$ and
$c_+\simeq c_-$, $a\simeq b\simeq 1/2$ and $\gamma \ll 1$, which
explains the time-evolutions in Fig. 4.

Here let us estimate $\Omega$ from $x({\pi\over \omega})$ and
$x({2\pi\over \omega})$. From Eq.(\ref{eqX2}) and  Eq.(\ref{eqL}),
\begin{equation}
x({\pi\over \omega})=p,
\label{eq22}
\end{equation}
and
\begin{equation}
x({2\pi\over \omega})=1-2p(1-\cos\alpha)+O(p^2).
\label{eq23}
\end{equation}
Thus
\begin{equation}
R={1-x({2\pi\over \omega})\over x({\pi\over \omega})}=2(1-\cos\alpha).
\label{eqcos}
\end{equation}
In Fig. 6 we show $R$ estimated from numerical calculations of
$x(\pi/\omega)$ and $x(2\pi/\omega)$ for various values of $H_0$,
which agrees with the $H_0$ dependence of $\Omega$ shown in Fig.5
($\Omega=\sqrt{pR}\times \omega/2\pi$). In Fig. 5,
$\Omega_{\rm  max}=\sqrt{p}\times \omega/\pi$ is shown by a dotted line.
We find that $\sqrt{p}\times\omega/\pi$ gives the envelope of $\Omega(H_0)$
and confirm that the relation Eq.(\ref{eqtanO2}) holds.

Generally we know neither $\alpha$ nor $p$. Even in such situation,
we can estimate $\Omega_{\rm max}$ by
observing $\Omega$ for various $H_0$ and $\omega$.
Alternately, from $\Omega_{\rm max}$, we can estimate $p$ by
\begin{equation}
\Omega_{\rm max} = {2\omega\sqrt{p}\over\pi},
\label{eqOmax}
\end{equation}
and therefore also  $Delta E$ by making use of the relation Eq.(\ref{eqp}).
The present analysis is good for any periodic function of $H(t)$,
not necessarily $\cos(\omega t)$. We have confirmed the same behavior for a
piecewise linear function (i.e., zigzag function) of $H(t)$.

The present oscillation of $M(t)$ is due to the non-adiabatic transition
and is a peculiar property of quantum dynamics with time dependent field.
The present mechanism is so simple that it would be applicable for many cases
where the non-adiabatic transition takes place and that we hope
such nontrivial oscillation would be observed in an experiment of
nanoscale systems.

The present study is partially supported by the Grant-in-Aid for Scientific
Research from the Ministry of Education, Science and Culture.

\begin{figure}
\caption{Fig.1 Energy levels for $\gamma = 0.5$.}
\label{fig1}
\end{figure}

\begin{figure}
\caption{Fig.2 Time dependence of magnetization, $M(t)$ is
shown by a solid line. $H(t)+2.5$ is shown by a dotted line.}
\label{fig2}
\end{figure}

\begin{figure}
\caption{Fig.3 Nontrivial oscillation of magnetization $M(t)$.
$x(t)$ and $H(t)$ are also shown. }
\label{fig3}
\end{figure}

\begin{figure}
\caption{(a) Time dependences of magnetization for various
amplitudes, $H_0$.
(b) Time dependences of magnetization for various
frequencies $\omega$.}
\label{fig4b}
\end{figure}

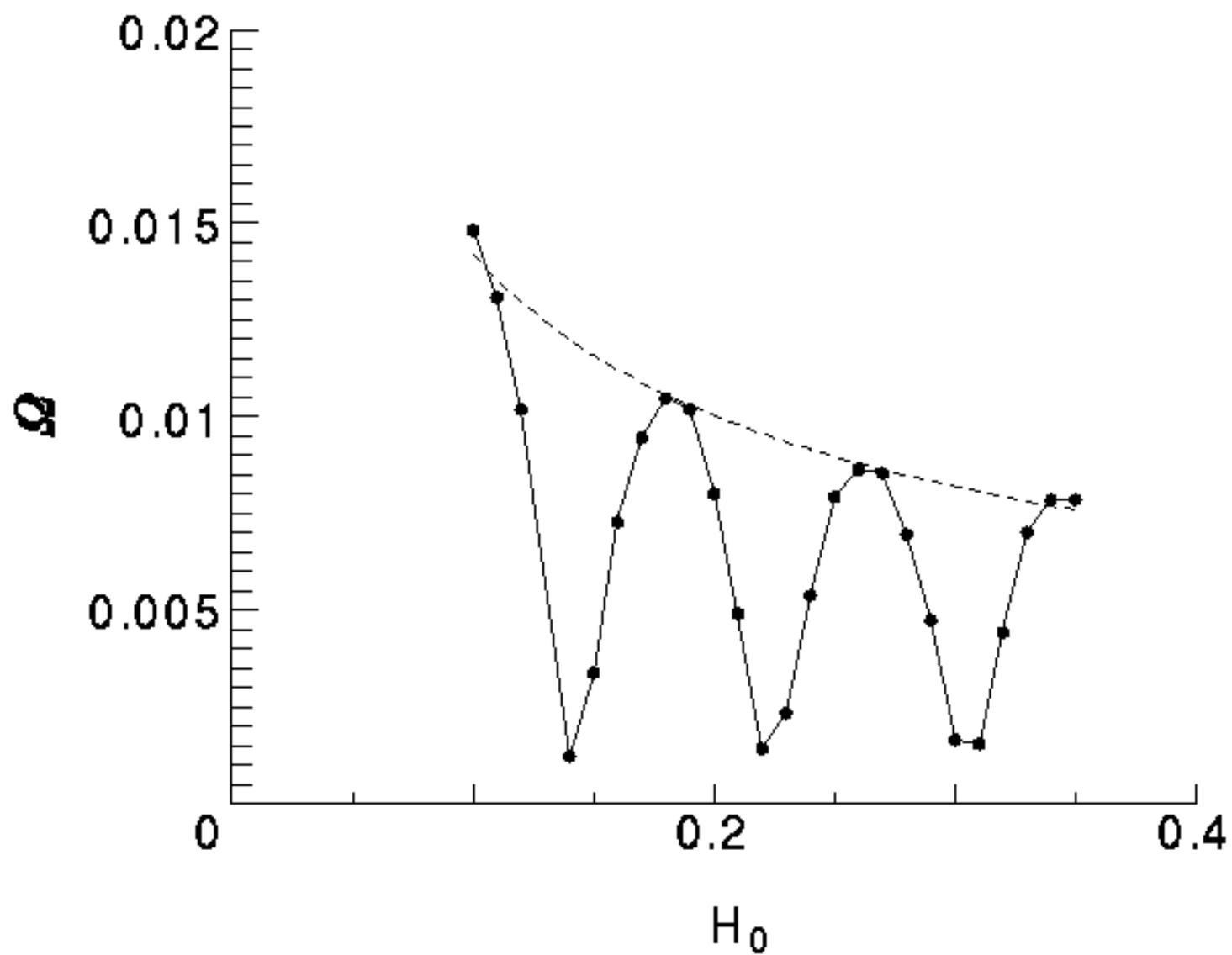
\begin{figure}
\caption{$H_0$ dependence of $\Omega$.
$\Omega_{\rm max}$ is shown by a dotted line. }
\label{fig5}
\end{figure}

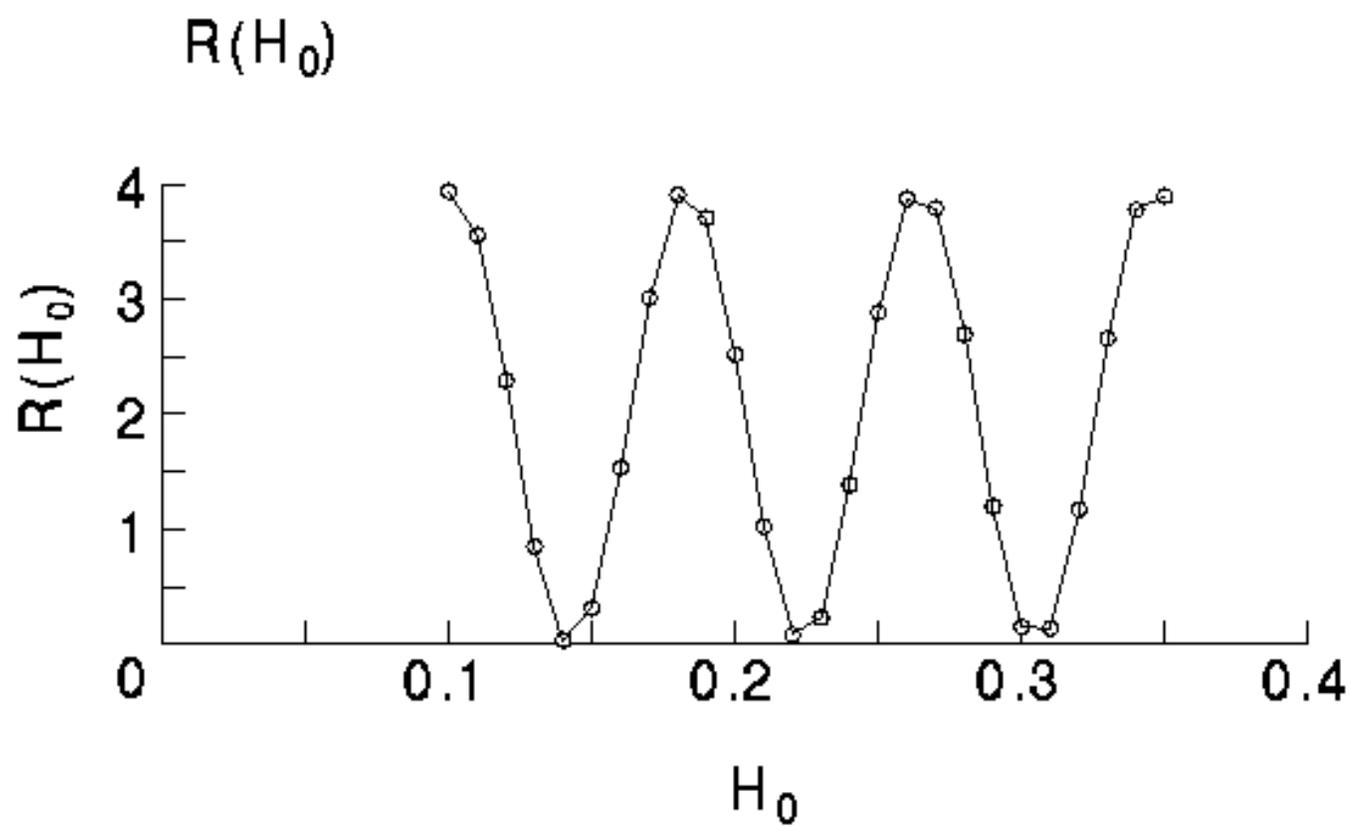
\begin{figure}
\caption{$H_0$ dependence of $R$.}
\label{fig6}
\end{figure}


\begin{references}
\bibitem{thmy1} S. Miyashita, J. Phys. Soc. Jpn. {\bf 64}, 3207 (1995)
\bibitem{thmy2} S. Miyashita, J. Phys. Soc. Jpn. {\bf 65}, 2734 (1996)
\bibitem{thmy3} H. De Raedt, S. Miyashita, K. Saito, D. Garc\'ia-Pablos,
 and N.Garc\'ia, unpublished.
\bibitem{th4} D. Garc\'ia-Pablos, N.Garc\'ia, P. Serena, and H. De Raedt,
Phys. Rev. B{\bf 53}, 741 (1996)
\bibitem{th5} D. Garc\'ia-Pablos, N.Garc\'ia, and H. De Raedt,
Phys. Rev. B{\bf 55}, 931 (1997)
\bibitem{th6} D. Garc\'ia-Pablos, N.Garc\'ia, and H. De Raedt,
Phys. Rev. B{\bf 55}, 937 (1997)
\bibitem{th7} E. M. Chudnovsky, Science {\bf 274}, 938 (1996)
\bibitem{th8} P. C. E. Stamp, Nature {\bf 383}, 125 (1996)
\bibitem{exp1} J. R. Friedman and M. P. Sarachik, T. Tejada and
R. Ziolo, Phys. Rev. Lett. {\bf 76},
3830 (1996)
\bibitem{exp2} L. Thomas, F. Lionti, R. Ballou, D. Gatteschi,
R. Sessoli and B. Barbara, Nature {\bf 383}, 145 (1996)
\bibitem{exp3} J. M. Hernandez, X. X. Zhang, F. Luis, and T. Tejada,
J. R. Friedman, M. P. Sarachik and R. Ziolo,
Phys. Rev. B {\bf 55}, 5858 (1997)
\bibitem{exp4} L. Thomas et al., Nature {\bf 383}, 145 (1996)
\bibitem{exp5} S. Gider, D. D. Awschalom, T. Douglas, S. Mann and
M. Chaparala, Science {\bf 268}, 77 (1995)
\bibitem{Landau} L. Landau, Phys. Z. Sowjetunion {\bf 2}, 46 (1932)
\bibitem{Zener} C. Zener, Proc. R. Soc. London, Ser. A{\bf 137}, 696 (1932)
\bibitem{St} E. C. G. St\"uckelberg, Helv. Phys. Acta {\bf 5}, 369 (1932)
\bibitem{Pd} J.-M. Lopez-Castillo, A. Filali-Mouhim and J.-P. Jay-Gerin,
J. Chem. Phys. {\bf 97}, 1905 (1992)
\bibitem{Suzuki1} M. Suzuki, J. Phys. Soc. Jpn. {\bf 61}, 3015 (1992)
\bibitem{Suzuki2} M. Suzuki, Proc. Japan. Acad. Ser. B {\bf 69}, 161 (1993)
\end{references}
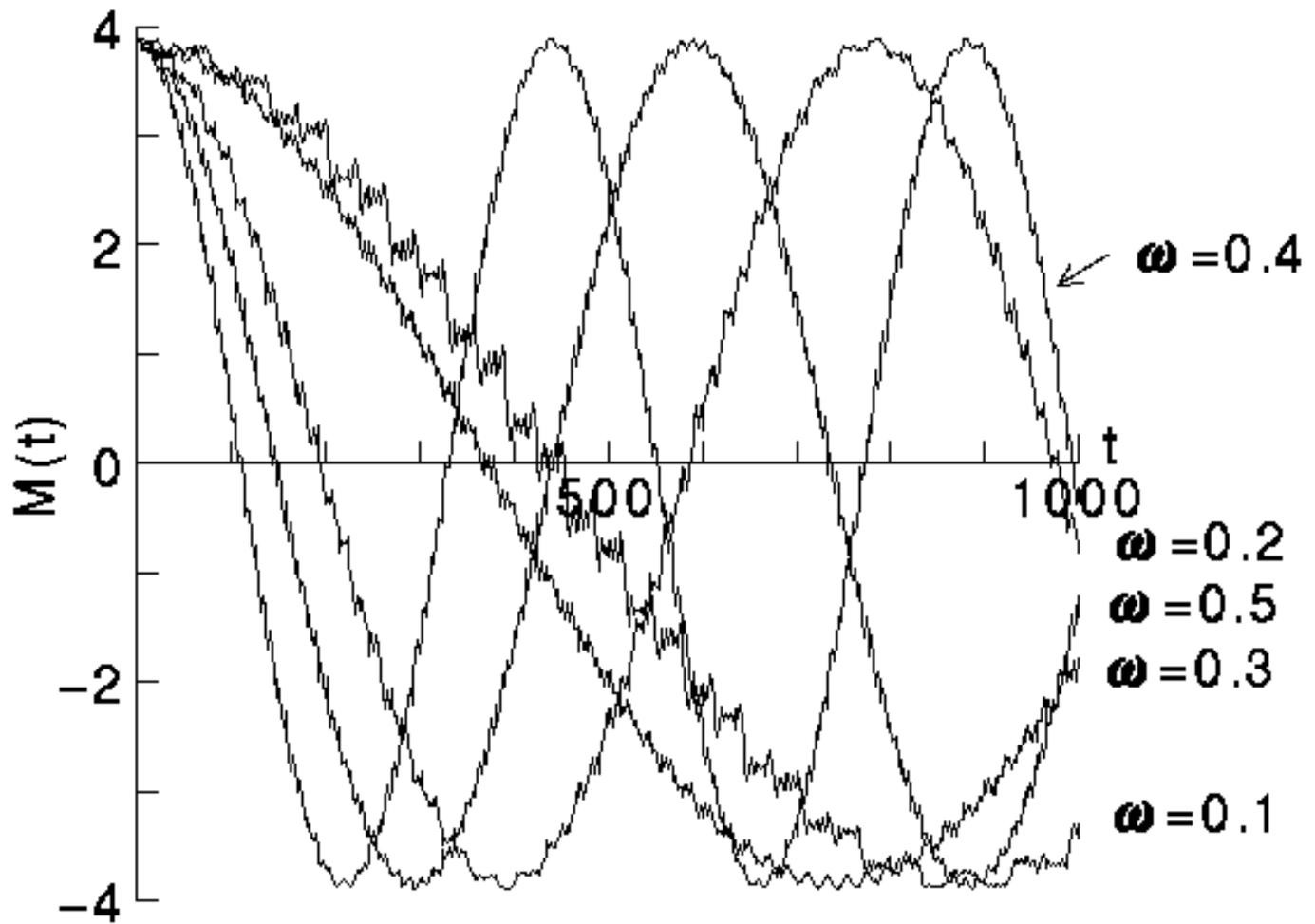
\end{document}